\documentclass[runningheads,openany]{svmult}

\usepackage{makeidx}   
\usepackage{graphicx}  
\usepackage{subeqnar}  
\usepackage{multicol}  
\usepackage{physmubb}  
\usepackage{rotating,epsfig} 
\usepackage{bbm}       
\usepackage{bm}        
\usepackage{url}       
\usepackage[OT2,OT1]{fontenc}
\usepackage{floatflt}
\usepackage{pstricks}
\usepackage{amsmath}
\usepackage{amssymb}
\usepackage{slashed}
\usepackage{macros}

\usepackage{harvard}
\harvardparenthesis{square}

\usepackage{booktabs}  
\usepackage{dcolumn}
\newcolumntype{d}{D{.}{.}{-1}}
\makeindex             

\begin{document}

\frontmatter



\mainmatter


\title{Real-Time Dynamics and Conical Intersections}
\author{M.E. Casida, B. Natarajan, and T. Deutsch}
\authorrunning{M.\protect\,E.\protect\,Casida, B.\protect\, Natarajan, and T.\protect\, Deutsch}
\tocauthor{M.\protect\,E.\protect\,Casida, B.\protect\, Natarajan, and T.\protect\, Deutsch}

\maketitle
\label{chap_4.1}

\textbf{Mark E.\ Casida} \\
{\em Laboratoire de Chimie Th\'eorique},\\
{\em D\'epartement de Chimie Mol\'ecularie (DCM, UMR CNRS/UJF 5250)},\\
{\em  Institut de Chimie Mol\'eculaire de Grenoble (ICMG, FR2607)},\\
{\em  Universit\'e Joseph Fourier (Grenoble I)},\\
{\em 301 rue de la Chimie, BP 53},\\
{\em F-38041 Grenoble Cedex 9, FRANCE}\\
\texttt{Mark.Casida@UJF-Grenoble.Fr}

\textbf{Bhaarathi Natarajan} \\
{\em Laboratoire de Chimie Th\'eorique},\\
{\em D\'epartement de Chimie Mol\'ecularie (DCM, UMR CNRS/UJF 5250)},\\
{\em  Institut de Chimie Mol\'eculaire de Grenoble (ICMG, FR2607)},\\
{\em  Universit\'e Joseph Fourier (Grenoble I)},\\
{\em 301 rue de la Chimie, BP 53},\\
{\em F-38041 Grenoble Cedex 9, FRANCE}\\
and\\
{\em CEA, INAC, SP2M, L\_Sim},\\
{\em 38054 Grenoble Cedex 9, FRANCE}\\
\texttt{bhaarathi.natarajan@UJF-Grenoble.FR}

\textbf{Thierry Deutsch} \\
{\em CEA, INAC, SP2M, L\_Sim},\\
{\em 38054 Grenoble Cedex 9, FRANCE}\\
\texttt{thierry.deutsch@cea.fr}

\hspace{1cm}

\section{Introduction}
\label{sec:intro_4.1}

The area of excited-state dynamics is receiving increasing attention
for a number of reasons, including the importance of photochemical
processes in basic energy sciences, improved theoretical methods
and the associated theoretical understanding of photochemical
processes, and the advent of femtosecond (and now attosecond) spectroscopy allowing
access to more detailed experimental information about photochemical
processes.  Since photophysical and chemical processes are more complex than
thermal (i.e., ground state) processes, simulations quickly become
expensive and even unmanageable as the model system becomes 
increasingly realistic.  With its combination of simplicity and
yet relatively good accuracy, TDDFT has been finding an increasingly
important role to play in this rapidly developing field.  After 
reviewing some basic ideas from photophysics and photochemistry, this chapter
will cover some of the strengths and weaknesses of TDDFT for
modeling photoprocesses.  The emphasis will be on going beyond the
Born-Oppenheimer approximation.

There are distinct differences between how solid-state physicists and
chemical physicists view photoprocesses.  We believe that some of this
is due to fundamental differences in the underlying phenomena being
studied but that much is due to the use of different approximations
and the associated language.  Ultimately anyone who wants to work at
the nanointerface between molecules and solids must come to terms 
with these differences, but that is not our objective here.  Instead
we will adapt the point of view of a chemical physicist (or physical
chemist)---see e.g., Ref.~\cite{MB90}. 

\begin{figure}[t]
  \begin{center}
    \includegraphics[angle=0,width=1.0\textwidth]{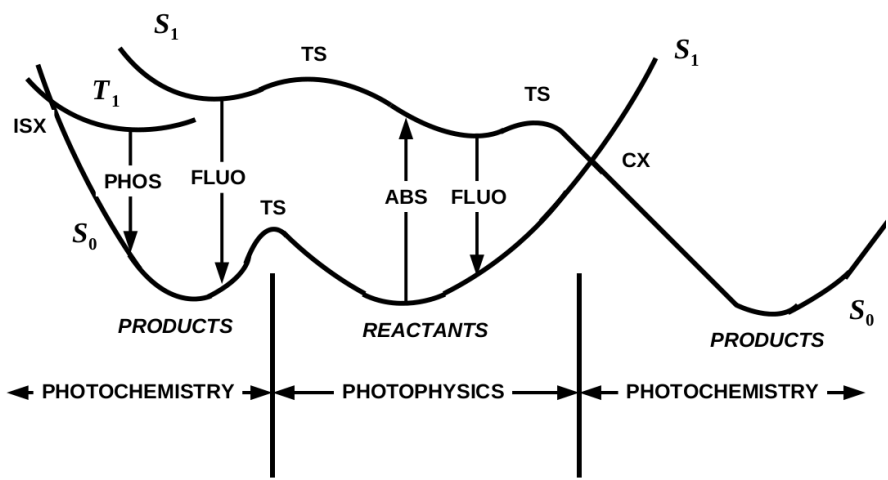}
  \end{center}
  \caption{\label{fig:PES}
    Schematic representation of potential energy surfaces for photophysical and photochemical
    processes: $S_0$, ground singlet state; $S_1$, lowest excited singlet state; $T_1$, lowest
    triplet state; ABS, absorption; FLUO, fluorescence; PHOS, phosphorescence; ISX, intersystem
    crossing; CX, conical intersection; TS, transition state.}
\end{figure}

The usual way to think about molecular dynamics is in terms of the
potential energy surfaces that come out of the Born-Oppenheimer
separation.  In thermal processes, vibrations are associated with
small motions around potential energy surface minima.  Chemical reactions are usually
described as going over passes (transition states) on these 
hypersurfaces as the system moves from one valley (reactants)
to another (products).  Photoprocesses are much more complicated
(Fig.~\ref{fig:PES}).  Traditionally they include not only processs
that begin by absorption of a photon, but also any process
involving electronically excited states, such as chemiluminescence
(e.g., in fireflies and glow worms)
where the initial excitation energy is provided by a chemical
reaction.  The Frank-Condon approximation tells us
that the initial absorption of a photon will take us from one
vibronic state to another without much change of molecular 
geometry, thus defining a Frank-Condon region on the excited-state
potential energy surfaces.  The molecule can return to the ground state by emitting
a photon of the initial wave length or, depending upon vibronic
coupling and perturbations from surrounding molecules, the molecule
may undergo radiationless relaxation to a lower energy excited
state before emitting or it may even decay all the way to the
ground state without emitting.  If emission takes place from
a long-lived excited state of the same spin as the ground state,
then we speak of fluorescence.  If emission takes place
from an excited-state with a different spin due to intersystem
crossing, then we speak of phosphorescence.  
If it is unsure whether the emission is fluorescence or phosphorescence, then we just say
the molecule luminesces.  Because of the large variety of de-excitation
processes, excited molecules usually return too quickly to their
ground state for the molecular geometry to change much.  We then
speak of a photophysical process because no chemical reaction 
has taken place.  Thus fluorescence is usually described as an excited
molecule relaxing slightly to a nearby minimum on the excited-state
potential energy surface where it is momentarily trapped before it emits to the ground
state.  It follows that the photon emitted during fluorescence is Stokes
shifted to a lower energy than the photon initially absorbed.

Photochemical reactions occur when the excited molecule decays to 
a new minimum on the ground state surface, leading to a new chemical
species (product).  This may have positive value as a way for synthesizing 
new mole-cules or negative value because of photodegradation of materials
or because of photochemically-induced cancers.  Either way the photochemical
reaction must occur quickly enough that it can compete with other decay
processes.  Photochemical reactions almost
always occur via photochemical funnels where excited-state and ground-state
surfaces come together, either almost touching (avoided crossing) or
crossing (conical intersection).  These funnels play a role in photochemical
reactions similar to transition states for thermal reactions.  However it must be kept
in mind that these funnels may be far from the Frank-Condon region on the
excited-state potential energy surface, either because there is an easy energetically-``downhill''
process or because, unless the absorption wavelength can be carefully tuned to a known
vertical excitation energy, the system will typically arrive in
an electronically-excited state with excess dynamical energy
which can be used to move from one excited-state  potentiall energy surface valley over a transition state
to funnels in another basin of the excited-state potential energy surface.  While conical intersections are forbidden
in diatomic molecules, they are now believed to be omnipresent in the photochemistry
of polyatomic molecules where the presence of an avoided crossing in potential energy surface cross-sections obtained
from oversimplified models (e.g., those making simplifying symmetry assumptions) usually
indicates a nearby conical intersection.  
A particularly striking example is provided by experimental and theoretical evidence
that the fundamental photochemical reaction involved in vision passes through a conical 
intersection \cite{PAW+10}.
For these reasons, modern photochemical modeling often 
involves some type of dynamics and, when this is not possible, at least focuses on 
finding conical intersections that can explain the reaction.

While a single-reference electronic structure method may be adequate for describing photophysical 
processes, the usual standard for describing photochemical processes is a multireference 
electronic structure method such as the complete active space self-consistent field (CASSCF)
method.  (See Ref.~\cite{HJO00} for a review of modern quantum chemical methods.)
This is because the first approximation to the wave function along a reaction pathway is
as a linear combination of the wave functions of the initial reactants and the final products.
Since CASSCF is both computationally heavy and requires a high-level of user intervention, a 
simpler method such as TDDFT would be very much welcome.  Early work in TDDFT in quantum
chemistry foresaw increasing applications of TDDFT in photochemical modeling.  For example,
avoided crossings between cross-sections of excited-state potential energy surfaces may be described with TDDFT because of the
multireference-like nature of TDDFT excited states \cite{CCS98}.  However great attention must
also be paid to problems arising from the use of approximate functionals \cite{C02}.
In particular, the TDDFT Tamm-Dancoff approximation (TDA) \cite{HH99} was found to give
improved shapes of excited-state potential energy surfaces \cite{CGG+00,CJI+07}, albeit at the price of loosing 
oscillator strength sum rules.
A major advance towards serious investigations of TDDFT for describing photoprocesses
came with the implementation of analytical derivatives for photochemical excited states in
many electronic structure programs \cite{VA99,VA00,FA02,H03,RF05,DK05,SFM+06}.  This made
it possible to relax excited-state geometries and to calculate Stokes shifts within the
framework of TDDFT.  In fact, TDDFT has become a standard part of the photochemical modeler's
toolbox.  It is typically used for calculating absorption spectra and exploring excited-state
potential energy surfaces around the Frank-Condon region.  TDDFT also serves as a rapid way to gain the chemical
information needed to carry out subsequent CASSCF calculations.
(See e.g., Refs.~\cite{DZK01a,DZK01b,DKSZ02,SDK+02,DZ03} for some 
combined femtosecond spectroscopy/theoretical
studies of photochemical reactions which make good use of TDDFT.)
It would be nice to be able to use a single method to model entire photochemical 
processes.  The advent of mixed TDDFT/classical surface-hopping Tully-type dynamics
\cite{TTR07,WMSB08,TTR+08,TTR09,TTR09a,BPP+10} is giving us a way to extend the power
of TDDFT to the exploration of increasingly complicated photochemical processes.

The rest of this chapter is organized as follows: The next section reviews non-Born-Oppenheimer phenomena
from a wave-function point of view, with an emphasis on mixed quantum/classical dynamics.
This sets the stage for our discussion of TDDFT for real-time dynamics and conical intersections in 
Sec.~\ref{sec:TDDFT_4.1}.  We sum up in Sec.~\ref{sec:conclude_4.1}.  

\section{Wave-Function Theory}
\label{sec:WF_4.1}

Most likely anyone who has made it this far into this chapter has seen the Born-Oppenheimer approximation
at least once, if not many times.  However it is relatively rare to find good discussions
that go beyond the Born-Oppenheimer approximation~\cite{DM02,C04}.  This section tries to 
do just this from a wave-function point of view, in preparation for a discussion of TDDFT 
approaches to the same problems in the following section.  We first begin by reviewing 
(again!) the Born-Oppenheimer approximation, but this time with the point of view of identifying the missing 
terms.  We then discuss mixed quantum/classical approximations, and end with a discussion 
of the pathway method and ways to find and characterize conical intersections.  {\em We shall use Hartree atomic
units ($\hbar = m_e = e = 1$) thoughout and adapt the convention in this section that electronic 
states are labeled by small Latin letters, while nuclear degrees of freedom are labeled by 
capital Latin letters.}

\paragraph{Born-Oppenheimer Approximation and Beyond}$\mbox{ }$
\label{subsec:BO}

As is well-known, the Born-Oppenheimer approximation relies on a separation of time scales: Since electrons
are so much lighter and so move so much faster than nuclei, the electrons may be thought of as 
moving in the field of nuclei which are ``clamped'' in place and the nuclei move in a field which 
is determined by the mean field of the electrons.  The Born-Oppenheimer approximation provides a precise 
mathematical formulation of this physical picture.  Our interest here is in where the Born-Oppenheimer approximation 
breaks down and what terms are needed to describe this breakdown.  

Consider a molecule composed of $M$ nuclei and $N$ electrons.  Denote the nuclear coordinates
by ${\bf R} = ({\vec R}_1, {\vec R}_2, \ldots , {\vec R}_M)$ and electronic coordinates
by ${\bf r} = ({\vec r}_1, {\vec r}_2, \ldots , {\vec r}_N)$.  The full Hamiltonian, 
$ {\hat H}({\bf R},{\bf r}) = {\hat T}_n({\bf R}) + {\hat H}_e({\bf r};{\bf R}) + V_{nn}({\bf R}) $, 
is the sum of an electronic Hamiltonian,
${\hat H}_e({\bf r};{\bf R}) = {\hat T}_e({\bf r}) + V_{en}({\bf r};{\bf R}) + V_{ee}({\bf r}) $,
with its electron kinetic energy, ${\hat T}_e$, electron-nuclear attraction, $V_{en}$, and electron-electron
repulsion, $V_{ee}$, with the missing nuclear terms---namely the nuclear kinetic energy, $\hat{T}_n$, and the
nuclear-nuclear repulsion, $V_{nn}$.  Solving the time-dependent Schr\"odinger equation,
\begin{equation}
  {\hat H}({\bf R},{\bf r}) \Phi({\bf R},{\bf x},t) = i\frac{d }{dt } \Phi({\bf  R},{\bf x},t) 
 \, ,
  \label{eq:BO_4.1_1}
\end{equation}
is a formidable $(N+M)$-body problem.  
(${\bf x}$ denotes inclusion of electron spin.  We have decided to omit nuclear spin, though
this should be included in principle when identical nuclei with spin are present, such as in the
case of {\em ortho}- and {\em para}-hydrogen.)
That is why the Born-Oppenheimer expansion (which is not yet the Born-Oppenheimer approximation!),
\begin{equation}
  \Phi({\bf R},{\bf x},t) = \sum_j \Psi_j({\bf x};{\bf R}) \chi_j({\bf R},t) \, ,
  \label{eq:BO_4.1_2}
\end{equation}
is used,
where the electronic wave functions are solutions of the time-independent electronic problem in the field
of clamped nuclei,
${\hat H}_e({\bf r};{\bf R}) \Psi_j({\bf x};{\bf R}) = \\ E_j^e({\bf R}) \Psi_j({\bf x};{\bf R})$.
Inserting the Born-Oppenheimer expansion [Eq.~(\ref{eq:BO_4.1_2})] into the full Schr\"odinger equation 
[Eq.~(\ref{eq:BO_4.1_1})], left multiplying by $\Psi_i^*({\bf x};{\bf R})$, and
integrating over ${\bf x}$ gives the time-dependent Schr\"odinger equation for the nuclear
degrees of freedom,
\begin{equation}
  \left( {\hat T}_n({\bf R}) + V_i({\bf R}) \right) \chi_i({\bf R},t) 
  + \sum_j {\hat V}_{i,j}({\bf R}) \chi_j({\bf R},t) 
  = i \frac{\partial}{\partial t} \chi_i({\bf R},t)
  \, .
  \label{eq:BO_4.1_3}
\end{equation}
Here, $V_i({\bf R}) = E_i^e({\bf R}) + V_{nn}({\bf R})$, is the {\em adiabatic} potential energy
surface for the $i$th 
electronic state.  [Notice that this is a different use of the term ``adiabatic'' than in the TDDFT 
``adiabatic approximation'' for the exchange-correlation (xc) functional.]
The remaining part, $\hat{V}_{i,j}({\bf R})$, is
the hopping term which couples the $I$th and $J$th PESs together.  As is well known, the Born-Oppenheimer 
approximation neglects the hopping terms,  
$\left( {\hat T}_n({\bf R}) + V_i({\bf R}) \right)\\ \chi_i({\bf R},t) = 
i (\partial/\partial t) \chi_i({\bf R},t)$.

We, on the other hand, are interested in precisely the terms neglected by the Born-Oppenheimer approximation.
The hopping term is given by,
${\hat V}_{i,j}({\bf R}) \chi_j({\bf R},t) =
  - \sum_I (1/2m_I) \left( G^{(I)}_{i,j}({\bf R})
  + 2 {\vec F}_{i,j}^{(I)}({\bf R}) \cdot {\vec \nabla}_I \right) \chi_j({\bf R},t)$,
where,\\
$G_{i,j}^{(I)}({\bf R})
  = \int \Psi_i^*({\vec x};{\vec R}) \left( \nabla_I^2 \Psi_j({\vec x};{\bf R}) \right) \, d{\vec x}
  = \langle i \vert \nabla_I^2  \vert j \rangle$,
is the scalar coupling matrix and,
${\vec F}_{i,j}^{(I)}({\bf R})
  = \int \Psi_i^*({\vec x};{\bf R}) \left( {\vec \nabla}_I \Psi_j({\vec x};{\bf R}) \right) \, d{\vec x}
  = \langle i \vert {\vec \nabla}_I  \vert j \rangle$,
is the derivative coupling matrix \cite{C04}.  Note that the derivative coupling matrix is 
also often denoted ${\vec d}_{i,j}^{I}$ and called the nonadiabatic coupling  vector \cite{DM02}.
Here we have introduced a compact notation for some complicated objects: The scalar coupling matrix 
is simultaneously 
a function of the nuclear coordinates, a matrix in the electronic degrees of freedom, and a vector 
in the nuclear degrees of freedom, and a matrix in the electronic degrees of freedom.  The derivative coupling matrix is 
simultaneously a function of the nuclear coordinates, a matrix in the electronic degrees of freedom,
a vector in the nuclear degrees of freedom {\em and} a vector in the three spatial coordinates of
the $I$th nucleus.  

Interestingly the scalar coupling matrix and derivative coupling matrix are not independent objects.  Rather, making use of the resolution
of the identity for the electronic states, it is straightforward to show that,\\
$\sum_k \left( \delta_{i,k} {\vec \nabla}_I + {\vec F}^{(I)}_{i,k}(\bf R) \right) \cdot
  \left( \delta_{k,j} + {\vec \nabla}_I {\vec F}^{(I)}_{k,j}(\bf R) \right)
  = \nabla_I^2 + G^{(I)}_{i,j}(\bf R) +\\ 2 {\vec F}^{(I)}_{i,j}(\bf R) \cdot {\vec \nabla}_I$.
We may then rewrite the time-dependent nuclear equation~(\ref{eq:BO_4.1_3}) as,
\begin{eqnarray}
  & -& \left\{ \sum_I \frac{1}{2 m_I} 
  \left[ \sum_k \left( \delta_{i,j} {\vec \nabla}_I + {\vec F}^{(I)}_{i,k}(\bf R) \right) \cdot
  \left( \delta_{k,j} {\vec \nabla}_I + {\vec F}^{(I)}_{k,j}(\bf R) \right) \right] \right\} 
   \chi_j({\bf R},t)
  \nonumber \\
  & + & V_i({\bf R}) \chi_i({\bf R},t)
  = i \frac{\partial }{\partial t} \chi_j({\bf R},t) \, ,
  \label{eq:BO_4.1_8}
\end{eqnarray}
which is known as the group Born-Oppenheimer equation \cite{C04}.  Evidently this is an equation which can be
solved within a truncated manifold of a few electronic statees in order to find fully quantum
mechanical solutions beyond the Born-Oppenheimer approximation.

More importantly for present purposes is that Eq.~(\ref{eq:BO_4.1_8}) brings out the importance
of the derivative coupling matrix.  The derivative coupling matrix can be rewritten as,
\begin{equation}
  {\vec F}_{i,j}^{(I)}({\bf R}) = 
   \frac{\langle i \vert \left( {\vec \nabla}_I {\hat H}_e({\bf R}) \right) \vert j \rangle -
   \delta_{i,j}  {\vec \nabla}_I E_i^e({\bf R})  }{E_j^e({\bf R})-E_i^e({\bf R})} \, .
  \label{eq:BO_4.1_9}
\end{equation}
Since this equation is basically a force-like term,
divided by an energy difference, we see that we can neglect coupling between adiabatic potential
energy surfaces when (i) the force on the nuclei is sufficiently small (i.e., the nuclei
are not moving too quickly) and (ii) when the energy difference between potential energy surfaces 
is not too large.  These conditions often break down in funnel regions of photochemical reactions.
There is then a tendency to follow diabatic surfaces, which may be defined rigorously by
a unitary transformation of electronic states (when it exists) to a new representation satisfying
the condition, ${\vec F}_{i,j}^{(I)}(\bf R) \approx 0$.  The advantage of the diabatic
representation (when it  exists, which is not always the case) is that it eliminates the 
off-diagonal elements of the derivative coupling matrix in the group Born-Oppenheimer equation [Eq.~(\ref{eq:BO_4.1_8})], hence
eliminating the need to describe surface hopping.  At a more intuitive level,
the character of electronic states tends to be preserved along diabatic surfaces because
$\langle i \vert d j/d t \rangle = {\dot {\vec R}} \cdot \langle i \vert {\vec \nabla} j \rangle
= {\dot {\vec R}} \cdot {\vec F}_{i,j} \approx 0$ in this representation.  For this reason, 
it is usual to trace diabatic surfaces informally in funnel regions by analyzing electronic 
state character, rather than seeking to minimize the nonadiabatic coupling vector.  Avoided crossings of adiabatic surfaces 
are then described as due to configuration mixing of electronic configurations belonging 
to different diabatic surfaces.

\paragraph{Mixed Quantum/Classical Dynamics}$\mbox{ }$
\label{subsec:mixed}

Solving the fully quantum-mechanical dynamics problem of coupled electrons and nuclei is a challenge
for small molecules and intractable for larger molecules.  Instead it is usual to use
mixed quantum/classical methods in which the nuclei are described by Newtonian classical mechanics 
while the electrons are described by quantum mechanics.  Dividing any quantum system into two
parts and then approximating one using classical mechanics is the subject of on-going
research \cite{K06}.  In general, no rigorous derivation is possible  and wave-function
phase information (e.g., the Berry phase) is lost which may be important in some instances.
Nevertheless mixed quantum/classical approximations are intuitive: Most nuclei (except perhaps 
hydrogen) are heavy enough that tunneling and other quantum mechanical effects are minor,
so that classical dynamics is often an {\em a priori} reasonable first approximation.  
Of course, rather than thinking of a single classical trajectory for the nuclear degree of 
freedom, we must expect to think in terms of ensembles (or ``swarms'') of trajectories 
which are built to incorporate either finite temperature effects or to try to represent 
quantum mechanical probability distributions or both.  The purpose of this subsection is
to introduce some common mixed quantum/classical methods.

The most elementary mixed quantum/classical approximation is Ehrenfest dynamics.
According to Ehrenfest's theorem \cite{E27},  Newton's equations are satisfied for mean values
in quantum systems, $d\langle {\hat {\vec r}} \rangle /dt = \langle {\hat {\vec p}} \rangle / m$ and
$d\langle {\hat {\vec p}} \rangle/dt = -\langle {\vec \nabla} V \rangle$ .
Identifying the position of the nuclei with their mean value, we can then write an
equation,
$m_I {\ddot {\vec R}}_I(t) = - {\vec \nabla}_I V({\bf R}(t))$,
whose physical interpretation is that the nuclei are moving in the mean field of the electrons.
Here
\begin{equation}
   V({\bf R}(t)) = 
  \langle \Psi({\bf R},t) \vert {\hat H}_e({\bf R}(t))\vert  \Psi({\bf R},t) \rangle  
  + V_{nn}({\bf R}(t)) \, ,
   \label{eq:mixed_4.1_2}
\end{equation}
where the electronic wave function is found by solving the time-dependent equation,
${\hat H}_e({\bf x},{\bf R}(t)) \Psi({\bf x};{\bf R},t)
   = i (\partial/\partial t) \Psi({\bf x};{\bf R},t)$.
While Ehrenfest dynamics has been widely and often successfully applied, it suffers from
some important drawbacks.  The first drawback is that the nuclei always move on average
potential energy surfaces, rather than adiabatic or diabatic surfaces, even when far from
funnel regions where the nuclei would be expected to move on the surface of a single
electronic state.  While this is serious enough, since it suggests errors
in calculating branching ratios (i.e., relative yields of different products
in a photoreaction), a more serious drawback is a loss of microscopic reversibility.
That is, the temporal variation of the mean potential energy surface depends upon past history and can easily 
be different for forward and reverse processes.

A very much improved scheme is the fewest switches method of Tully \cite{T90,HT94}.
Here the nuclei move on well-defined adiabatic potential energy surfaces,
$m_I {\ddot {\vec R}}_I(t) = -{\vec \nabla}_I V_i ({\bf R}(t))$,
and the electrons move in the field of the moving nuclei,
${\hat H}_e({\bf r};{\bf R}(t)) \Psi({\bf r},t) = i (d/dt) \Psi({\bf r},t)$.
To determine the probability that a classical trajectory describing nuclear motion
hops from one electronic potential energy surface to another, we expand 
$\Psi({\bf r},t) =  \sum_m \Psi_m({\bf r};{\bf R}(t)) C_m(t)$,
in solutions of the time-independent Schr\"odinger equation,
${\hat H}({\bf r};{\bf R}(t)) \Psi_m({\bf r};{\bf R}(t)) = E_m({\bf R}(t)) \Psi_m({\bf r};{\bf R}(t))$.
The probability of finding the system on surface $m$ is then given by,
$P_m(t) = \vert C_m(t) \vert^2$.
The coefficients may be obtained in a dynamics calculation by integrating the first-order equation,
${\dot C}_m(t) =  -i E_m(t) C_m(t)  -  \sum_n \langle m \vert (d n/dt) \rangle \\ \times C_n(t)$.
A not unimportant detail is that the nonadiabatic coupling elements need not be calculated explicitly, but instead
can be calculated using the finite difference formula,
$\langle m(t+\Delta t/2) \vert {\dot n}(t+\Delta t/2) \rangle = \\
   \left[ \langle m(t) \vert n(t+\delta t) \rangle -  \langle m(t+\Delta t) \vert n(t) \rangle \right]  /(2\Delta t)$.
In practice, it is also important to minimize the number of surface hops or switches in order to 
keep the cost of the dynamics calculation manageable.  Tully accomplished this by introducing
his fewest-switches algorithm which is a type of Monte Carlo procedure designed to
correctly populate the different PESs with a minimum of surface hopping.  Briefly, the
probability of jumping from surface $m$ to surface $n$ in the interval $(t,t+\Delta t)$
is given by $g_{m \rightarrow n}(t,\Delta t) = {\dot P}_{m,n}(t)\Delta t /P_{m,m}(t)$ where
$P_{m,n}(t)=C_m(t)C_n^*(t)$.  A random number $\xi$ is generated with uniform probability
on the interval $(0,1)$ and compared with $g_{m \rightarrow n}(t,\Delta t)$.  The transition
$m \rightarrow n$ occurs only if $P_n^{(m-1)} < \xi < P_n^{(m)}$ where $P_n^{(m)} = \sum_{l=1,m} P_{n,l}$
is the sum of the transition probabilities for the first $m$ states.
Additional details of the algorithm, beyond the scope of this chapter, involve readjustment 
of nuclear kinetic energies and the fineness of the numerical integration grid for the 
electronic part of the calculation with respect to that of the grid for the nuclear degrees of freedom.

It is occasionally useful to have a simpler theory for calculating the probability of
potential energy surface hops which depends only on the potential energy surfaces and not on the wave functions.  Such a theory was suggested by 
Landau \cite{L32} and Zener \cite{Z32} (see also Wittig \cite{W05}).  Their work predates the modern appreciation 
of the importance of conical intersections and so focused on surface hopping at avoided crossings.  The Landau-Zener 
model assumes that surface hopping occurs only on the surface where the two diabatic 
surfaces cross that give rise to the avoided crossing where the surface hopping occurs.  After some 
linearizations and an asymptotic limit, it is possible to arrive at a very simple final formula,
\begin{equation}
   P = \exp \left( - \frac{\pi^2 \Delta E^2_{adia}}{h (d \vert \Delta E_{dia} \vert / dt) } \right)
   \, ,
   \label{eq:mixed_4.1_11}
\end{equation}
for the probability of hopping between two potential energy surfaces.  This formula is to be applied
at the point of closest approach of the two potential energy surfaces where the energy difference is
$\Delta E_{adia}$.  However $d \vert \Delta E_{dia} \vert / dt$
is evaluated as the maximum of the rate of change of the {\em adiabatic} energy difference 
as the avoided crossing is approached.  While not intended to be applied to conical intersections, it is still quite applicable in
photodynamics calculations since trajectories rarely go exactly through a conical intersection.

\paragraph{Pathway Method}$\mbox{ }$
\label{subsec:pathway}

Dynamics calculations provide a swarm of reaction trajectories.  The ``pathway method''
provides an alternative when dynamics calculations are too expensive or a simplified picture is 
otherwise desired, say, for interpretational reasons.  The pathway method consists of mapping 
out minimum energy pathways
between the initial Franck-Condon points obtained by vertical excitations and excited-state
minima or conical intersections.  Although analogous to the usual way of finding thermal reaction paths, it
is less likely to be a realistic representation of true photoprocesses except in the limit of threshold excitation
energies since excess energy is often enough to open up alternative pathways over excited-state
transition states.  While the necessary ingredients for the photochemical pathway method are similar to
those for thermal reactions, conical intersections are a new feature which is quite different from a thermal transition state.  This section provides a brief review for finding and characterizing conical intersections.

The notion of a conical intersection arises from a relatively simple argument \cite{Y01}.
The potential energy surface of a molecule with $f$ internal degrees of freedom is an $f$-dimensional hypersurface
in an $(f+1)$-dimensional space (the extra dimension is the energy axis).  If two potential energy surfaces simply
cross ``without seeing each other'', then the crossing space is characterized by the constraint,
\begin{equation}
   E_i({\bf R}) = E_j({\bf R}) \, ,
   \label{eq:pathway_4.1_1}
\end{equation}
makes the crossing space $(f-1)$-dimensional.  However in quantum mechanics, we also have the 
additional constraint,
\begin{equation}
   H_{i,j}({\bf R}) = 0 \, .
   \label{eq:pathway_4.1_2}
\end{equation}
This makes the crossing space $(f-2)$-dimensional.  This means that there will be two independent
directions in hyperspace in which the two potential energy surfaces will separate.  These two directions define a branching
plane.  Within the 3-dimensional space defined by the energy and the branching plane,
the conical intersection appears to be a double cone (Fig.~\ref{fig:cx1}), the point of which represents an entire $(f-1)$-dimensional 
space.  Of course, $f=1$ for a diatomic and no conical intersection is possible.  This is the origin of the well-known
avoided crossing rule for diatomics.  Here we are intererested in larger molecules where the low dimensionality
of the branching space in comparison with the dimensionality of the parent hyperspace can make
the conical intersection hard to locate and characterize.

In the pathway method, the system simply goes energetically downhill until two potential energy surfaces have the same
energy [Eq.~(\ref{eq:pathway_4.1_1})].  The resultant intersection space must be analyzed and
the branching plane extracted so that the surface crossing region can be properly visualized and interpretted.
In order to do so, let us recall a result from first-year calculus.  Imagine a trajectory, 
$\bf{R}(\tau)$, depending upon some parameter $\tau$ within the conical intersection surface.  Then 
${\vec \nabla} {\cal C}(\bf{R})$ must be perpendicular to the conical intersection
for any constraint function ${\cal C}(\bf{R})=0$ because,
$0 = d {\cal C}(\bf{R}(\tau))/d \tau
  = {\vec \nabla} {\cal C}(\bf{R}) \cdot (d \bf R/d \tau)$.
and we can always choose $d {\bf R}/d \tau \neq 0$.
Taking the gradient of Eq.~(\ref{eq:pathway_4.1_2}) defines the derivative coupling
vector,
${\vec f}_{i,j} = {\vec \nabla} H_{i,j}({\bf R})$,
while taking the gradient of Eq.~(\ref{eq:pathway_4.1_1}) defines the gradient difference vector,
${\vec g}_{i,j} = {\vec \nabla} E_i({\bf R}) - {\vec \nabla} E_j({\bf R})$.
Together the derivative coupling vector and gradient difference vector are referred to as the branching vectors which characterize the branching plane.
[Note that the derivative coupling vector is essentially the numerator of the derivative coupling matrix expression given in Eq.~(\ref{eq:BO_4.1_9}).
This confusion of nomenclature is unfortunate but present in the literature.]

These branching plane conditions are needed as constraints in the exploration of the conical intersection hyperspace when seeking
the minimum energy conical intersection or the first-order saddle point in conical interseection. In particular, 
there has been considerable effort devoted to the problem of developing efficient 
algorithms for finding minimum energy points within the conical intersection space \cite{KM85,AXR91,RRBO92,Y96,DS97,IK00}. 
Furthermore, an automated systematic exploration method for finding minimum energy conical 
intersections has very recently developed
\cite{MOM09}.  First-order saddle points and the corresponding minimum energy pathways 
both within the conical intersection 
hypersurface have been proposed to be important in dynamical 
trajectory simulations, and an optimization method was developed for such high-energy points
within the conical intersection hypersurface \cite{SBBR08}.
Some of the minimum energy conical intersection optimizers use the branching plane conditions explicitly to keep 
the degeneracy of the two adiabatic states during optimizations \cite{MY93,BRS94,AB97}, making
explicit use of both the derivative coupling vector and gradient difference vector at every step.
Most well-estabilished optimization algorithms assume smoothness of the 
function to be optimized.  Since the potential energy surface necessarily has a discontinuous
first derivative in the vicinity of a conical intersection, the above-mentioned algorithms for 
finding minimum energy conical intersections have required access to the gradient difference vector and derivative coupling  vectors.  The gradient difference vector can easily be obtained from
analytical gradients, if available, or by numerical energy differentiation if analytical
gradients are not yet available.  However methods for finding the derivative coupling vector are not yet
available for all methods since implementation of an analytical derivative method
is often regarded as a prerequisite \cite{CGP04,MOM10}.  Some approaches make use of
a penalty function to get around the need to calculate the derivative coupling vector and these have proven
very useful for finding minimum energy conical intersection regions without the need for the derivative coupling vector \cite{LCM08}.
This is especially important for methods such as renormalized coupled-cluster theories
and TDDFT or free-energy methods for which the electronic wave function is not completely 
defined, considerably complicating the problem of how to calculate derivative coupling vector matrix elements.
However, convergence of penalty function methods is in general slower than methods
which make explicit use of the branching plane constraints, especially if tight optimization of the
energy difference, $(E_i-E_j)$, is desired \cite{KKT07}. 

\section{TDDFT}
\label{sec:TDDFT_4.1}

The last section discussed the basic theory of non-Born-Oppenheimer dynamics and conical intersections from a wave-function point 
of view.  We now wish to see to what extent we can replace wave-function theory with what we hope 
will be a simpler DFT approach.  As usual in DFT, we seek both the guiding light of formal rigor 
and pragmatic approximations that work.  We will take a more or less historical approach to presenting 
this material.
{\em In this section, upper case Latin indices designate electronic states, while lower case Latin 
indices designate orbitals.}

One of the early objectives of TDDFT was to allow simulations of the behavior of atoms and clusters 
in intense laser fields, well beyond the linear-response regime and too complex to be handled by comparable 
wave-function methods.  The closely related topic of ion-cluster collisions was studied early on using 
TDDFT in a very simplified form \cite{YTAB98}.
The Ehrenfest method was the method of choice for TDDFT simulations coupling electronic and nuclear degrees
of freedom in this area. The main problem is how to take the expectation value in Eq.~(\ref{eq:mixed_4.1_2}).  This
is solved pragmatically by using,
$ V[\rho(t)]({\bf R}(t)) = T_s[\Phi_s(t)] +  \int \left( v_{en}({\bf r},{\bf R}(t)) +  v_{appl}({\bf r},t) \right)
   \rho({\bf r},t) \, d{\bf r} 
   +\\ (1/2)  \int \int \left( \rho({\bf r}_1,t)\rho({\bf r}_2,t) /r_{12} \right) d{\bf r}_1 d{\bf r}_2
   + E_{xc}[\rho(t)] + V_{nn}({\bf R}(t))$,
where $T_s$ is the usual Kohn-Sham noninteracting kinetic energy evaluated using the (now
time-dependent) Kohn-Sham determinant $\Phi_s$, $v_{en}$ is the electron-nuclear attraction potential,
$v_{appl}$ is the potential for any applied electric field, and $E_{xc}$ is the usual Hohenberg-Kohn-Sham
xc-energy.  Note that the presence of $E_{xc}$ is highly reminiscent of the TDDFT
adiabatic approximation that the xc-potential should react instantaneously and without memory to any
temporal change in the time-dependent charge density.  
Among the notable work done with this approximation is early studies of the dynamics of 
sodium clusters in intense laser fields \cite{CRS98}, the development of the time-dependent electron
localization function \cite{BMG05}, and (more recently) the study of 
electron-ion dynamics in molecules under intense laser pulses \cite{KNY09}.
Besides limitations associated with the TDDFT adiabatic approximation,
the TDDFT Ehrenfest method suffers from the same intrinsic problems as its wave-function brother---namely
that it is implicitly based on an average potential energy surface and so does not provide state-specific information, and
also suffers from problems with microscopic irreversibility.

To our knowledge, the first DFT dynamics on a well-defined excited-state potential energy surface was not based upon
TDDFT but rather on the older multiplet sum method  of Ziegler, Rauk, and Baerends \cite{ZRB77,D94}.
This was the work of restricted open-shell Kohn-Sham (ROKS) formalism of 
Irmgard Frank {\em et al}.~\cite{FHMP97} who carried out Carr-Parinello dynamics for the open-shell
singlet excited state $^1(i,a)$  using the multiplet sum method energy expression, $E_s = 2E[\Phi_{i\uparrow}^{a\uparrow}] 
-E[\Phi_{i\downarrow}^{a\uparrow}]$, where $\Phi_{i\sigma}^{a\tau}$ is the Kohn-Sham determinant with the 
$i\sigma$ spin-orbital replaced with the $a\tau$ spin-orbital.  Such a formalism suffers from all the
formal difficulties of the multiplet sum method, namely that it is just a first-order estimate of the energy using
a symmetry-motivated zero-order guess for the excited-state wave function and assumes that DFT works
best for states which are well-described by single determinants.  Nevertheless appropriate 
use of the multiplet sum
method can yield results similar to TDDFT.  A recent application of this method is to the study of the
mechanism of the electrocylic ring opening of diphenyloxirane \cite{FF09}.

The implementation of TDDFT excited-state derivatives in a wide variety of programs
not only means that excited-state geometry
optimizations may be implemented, allowing the calculation of the Stokes shift between
absorption and fluorescence spectra, 
but that the pathway method can be implemented to search for
conical intersections in TDDFT.  Unless nonadiabatic coupling matrix elements can be calculated within TDDFT ({\em vide infra}),
then a penalty method should be employed as described in the previous section under the pathway method.
This has been done by Levine, Ko, Quenneville, and Martinez using conventional TDDFT \cite{LKQM06}
and by Minezawa and Gordon using spin-flip TDDFT \cite{MG09}.  We will come back
to these calculations later in this section.

Thee most recent approach to DFT dynamics on a well-defined excited-state potential energy surface
is Tully-type dynamics \cite{T90,HT94,T98} applied within a mixed TDDFT/classical trajectory 
surface-hopping approach.  Surface-\\hopping probabilities can be calculated from potential energy surfaces alone within the  Landau-Zener 
method [Eq.~(\ref{eq:mixed_4.1_11})], however a strict application of Tully's method requires nonadiabatic coupling matrix elements as 
input.  Thus a key problem to be addressed is how to calculate nonadiabatic coupling matrix elements within TDDFT.
Initial work by Craig, Duncan, and Prezhdo used a simple approximation which neglected the
xc-kernel \cite{CDP05}.  A further approximation, criticized by Maitra \cite{M06}, has been made by 
Craig and co-workers \cite{CDP05,HCP06} who treated the electronic states as determinants of Kohn-Sham 
orbitals which are propagated according to the time-dependennt Kohn-Sham equation.  This means that 
neither the excitation energies nor the associated forces could be considered to be accurate.
The first complete mixed TDDFT/classical trajectory surface-hopping photodynamics method was proposed and implemented
by Tapavicza, Tavernelli, and R\"othlisberger \cite{TTR07} in a development version of the {\sc CPMD} code.
It was proposed that the nonadiabatic coupling matrix elements be evaluated within Casida's {\em ansatz} \cite{C95} which was originally
intended to aid with the problem of assigning excited states by considering a specific functional form for
an approximate excited-state wave function.  For the TDA, the Casida {\em ansatz} takes the familiar form,
$\Psi_I = \sum_{ia\sigma} \Phi_{i\sigma}^{a\sigma} X_{ia\sigma}$.
In fact, matrix elements between ground and excited states may be calculated exactly in a Casida-like formalism
because of the response theory nature of Eq.~(\ref{eq:BO_4.1_9}) \cite{CM00,HHS07,SF10}.  
Test results show reasonable accuracy for nonadiabatic coupling matrix elements as long as conical intersections are not approached too closely
\cite{B02,HHS07,TTR09,TTR09a,SF10}.  One likely reason for this is the divergence of Eq.~(\ref{eq:BO_4.1_9}) when $E_I=E_J$.
Hu and Sugino attempted to further improve the accuracy of nonadiabatic coupling  matrix elements by using average excitation
energies \cite{HS07}.  The problem of calculating nonadiabatic coupling matrix elements between two excited states is an open problem
in TDDFT, though the ability to calculated excited-state densities \cite{FA02} suggests that such matrix elements
could be calculated from double response theory using Eq.~(\ref{eq:BO_4.1_9}).
Soon after the implementation of mixed TDDFT/classical trajectory surface-hopping photodynamics in {\sc CPMD}, a very similar
method was implemented in {\sc TurboMol} and applied \cite{WMSB08,MWB08,BPP+10}.  
A version of {\sc TurboMol} capable of doing mixed TDDFT/classical trajectory surface-hopping photodynamics 
using analytic nonadiabatic coupling matrix elements has recently appeared \cite{SF10} and has been applied used to
study the photochemistry of vitamin-D \cite{TF11}.
Time-dependent density-functional tight-binding
(TDDFTB) may be regarded as the next step in a multiscale approach to the photodynamics of larger systems.  
From this point of view, it is interesting to note that mixed TDDFTB/classical trajectory surface-hopping photodynamics is
also a reality \cite{MWW+09}.  Given the increasingly wide-spread nature of implementations of mixed TDDFT classical 
trajectory surface-hopping photodynamics, we can only expect the method to be increasingly available to and used by the global
community of computational chemists.

\begin{figure}
  \begin{center}
    \includegraphics[angle=0,width=0.6\textwidth]{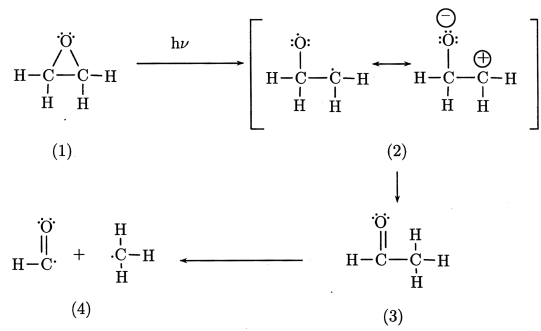}
  \end{center}
  \caption{\label{fig:GomerNoyes}
    Mechanism proposed by Gomer and Noyes in 1950 for the photochemical ring opening of
    oxirane.  Reprinted with permission from E.\ Tapvicza, I.\ Tavernelli, U.\ Rothlisberger, 
    C.\ Filippi, and M.\ E.\ Casida, {\em J. Chem. Phys.}
    {\bf 129}, 124108 (2008).  Copyright 2008, American Institute of Physics.
    }
\end{figure}

\begin{figure}
  \begin{center}
    \includegraphics[angle=0,width=1.0\textwidth]{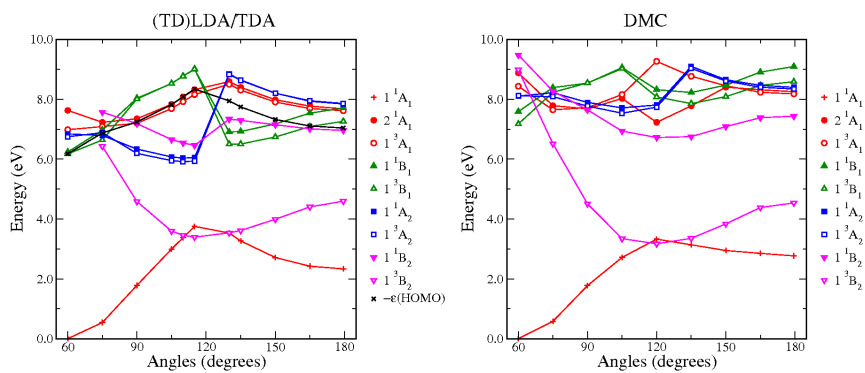}
  \end{center}
  \caption{\label{fig:C2v_1}
    Comparison of TDA TDLDA and diffusion Monte Carlo curves for $C_{2v}$ ring opening of oxirane.
    Reprinted with permission from F.\ Cordova, L.\ Joubert Doriol, A.\ Ipatov, M.\ E.\ Casida,
    C.\ Filippi, and A.\ Vela, {\em J. Chem. Phys.}
    {\bf 127}, 164111 (2007).  Copyright 2007, American Institute of Physics.
    }
\end{figure}

Before going further, let us illustrate the state-of-the-art for TDDFT when applied to non-Born-Oppenheimer dynamics and conical intersections.
We will take the example of the photochemical ring opening of oxirane (structure I in Fig.~\ref{fig:GomerNoyes}).
While this is not the ``sexy application''
modeling of some biochemical photoprocess, the photochemistry of oxiranes is not unimportant in synthetic photochemistry 
and, above all, this is a molecule where it was felt that TDDFT ``ought to work'' \cite{CJI+07}.
A first study showed that a main obstacle to photodynamics is the presence of triplet and near singlet
instabilities which lead to highly underestimated and even imaginary excitation energies as funnel regions 
are approached.  This is illustrated in Fig.~\ref{fig:C2v_1} 
for $C_{2v}$ ring opening.  While the real photochemical process involves asymetric CO ring-opening rather 
than the symmetric $C_{2v}$ CC ring-opening, results for the symmetric pathway have the advantage of being 
easier to analyze.  The figure shows that applying the TDA strongly attenuates the instability problem, putting
most curves in the right energy range.  Perhaps the best way to understand this is to realize that, whereas
time-dependent Hartree-Fock (TDHF), is a nonvariational method and hence allows variational collapse of excited
states, TDA TDHF is the same as configuration interaction singles (CIS) which is variational. 
There is however still a cusp in the ground state curve as the ground state configuration changes from
$\sigma^2$ to $(\sigma^*)^2$.  According to a traditional wave-function picture, these two states, which
are each double excitations relative to each other should be included in configuration mixing in order to
obtain a proper description of the ground state potential energy surface in the funnel region \cite{CJI+07,HNI+10}.

\begin{figure}
  \begin{center}
    \includegraphics[angle=0,width=0.5\textwidth]{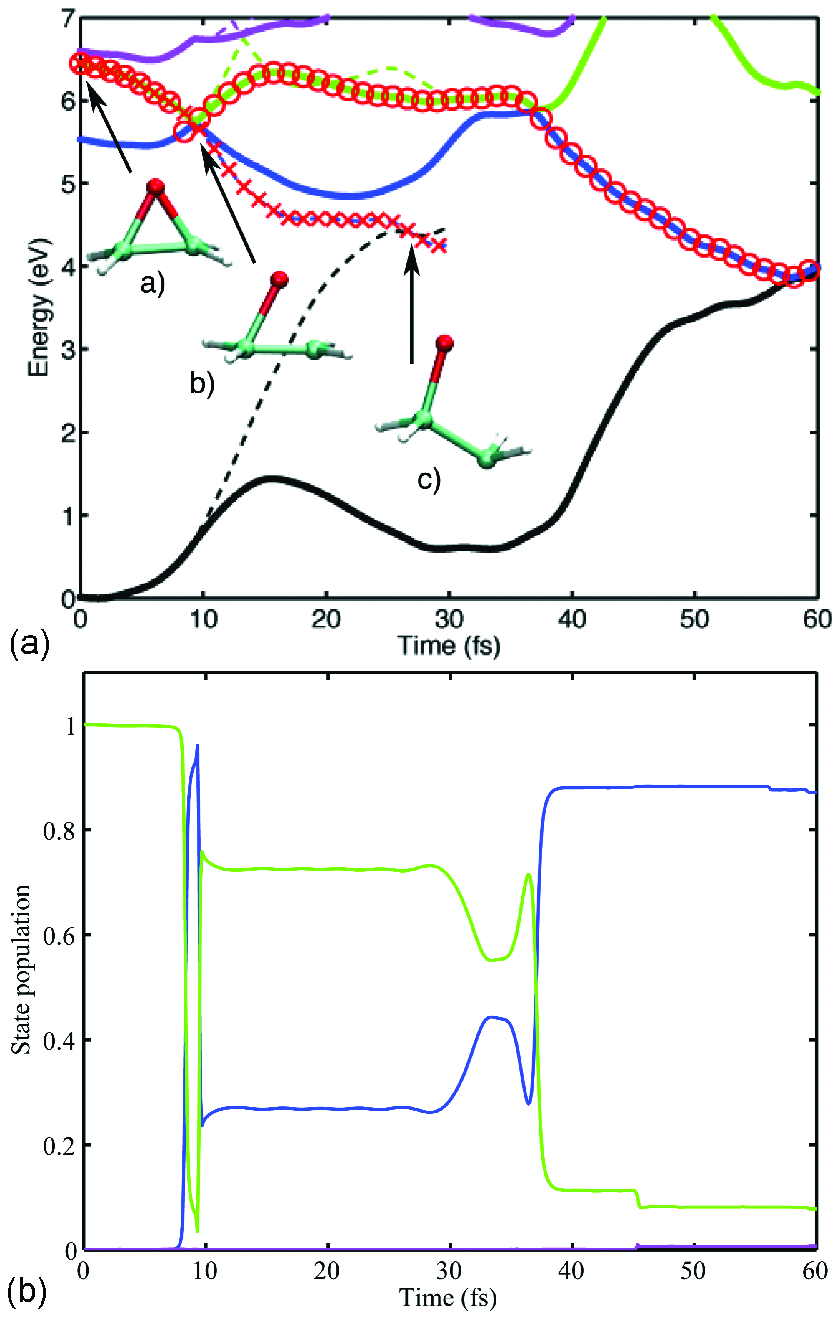}
  \end{center}
  \caption{\label{fig:trajectories}
  (a) Cut of potential energy surfaces along reaction path of an Landau-Zener (dashed line) and a fewest-switches (solid line) 
  trajectory (black, $S_0$; blue, $S_1$; green, $S_2$; magenta, $S_3$).  Both trajectories were started by 
  excitation into the $^1(n,3p_z)$ state, with the same geometry and same initial nuclear velocities. 
  The running states of the Landau-Zener and the fewest-switches trajectory are indicated by the red crosses and circles,
  respectively. The geometries of the Landau-Zener trajectory are shown at time a) 0, b) 10, and c) 30 fs.
  (b) State populations (black, $S_0$; blue, $S_1$; green, $S_2$; magenta, $S_3$) as a function of the
  fewest-switches trajectory in (a).
  Reprinted with permission from E.\ Tapvicza, I.\ Tavernelli, U.\ Rothlisberger, C.\ Filippi, and 
  M.\ E.\ Casida, {\em J. Chem. Phys.}
    {\bf 129}, 124108 (2008).  Copyright 2008, American Institute of Physics.
  }
\end{figure}

\begin{figure}
  \begin{center}
    \includegraphics[angle=0,width=0.5\textwidth]{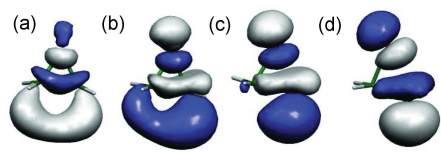}
  \end{center}
  \caption{\label{fig:orbitals}
  Change of character of the active state along the reactive Landau-Zener trajectory, shown in Fig.~\ref{fig:trajectories}. 
  Snapshots were taken at times (a) 2.6, (b) 7.4, (c) 12.2, and (d) 19.4 fs.  For (a) and (b), the running state is
  characerized by a transition from the highest occupied molecular orbital (HOMO) to the lowest unoccupied molecular
  orbital (LUMO) plus one (LUMO+1), while for (c) and (d) it is characterized by a HOMO-LUMO transition due to
  orbital crossing.  Note that the HOMO remains the same oxygen nonbonding orbital throughout the simulation.
  Reprinted with permission from E.\ Tapvicza, I.\ Tavernelli, U.\ Rothlisberger, C.\ Filippi, and
  M.\ E.\ Casida, {\em J. Chem. Phys.}
    {\bf 129}, 124108 (2008).  Copyright 2008, American Institute of Physics.
  }
\end{figure}

Figure~\ref{fig:trajectories} shows an example of mixed TDA TDPBE/classical trajectory surface-hopping calculations for 
the photochemical ring-opening of oxirane with the initial photoexcitation prepared in the $^1(n,3p_z)$ state.
Part (b) of the figure clearly shows that more than one potential energy surface is populated after about 10 fs.  The Landau-Zener
process is typical of the dominant physical process which involves an excitation from the HOMO nonbonding
lone pair on the oxygen initially to a $3p_z$ Rydberg orbital.  As the reaction proceeds, the ring opens
and the target Rydberg orbital rapidly changes character to become a CO $\sigma^*$ antibonding orbital
(Fig.~\ref{fig:orbitals}).  
Actual calculations were run on a swarm of 30 trajectories, confirming the mechanism previously proposed
Gomer-Noyes mechanism \cite{GN50} (Fig.~\ref{fig:GomerNoyes}), but also confirming other experimental 
by-products and giving unprecedented state-specific reaction details such as the orbital description
briefly described above.

\begin{figure}
   \begin{center}
   \includegraphics[angle=0,width=1.0\textwidth]{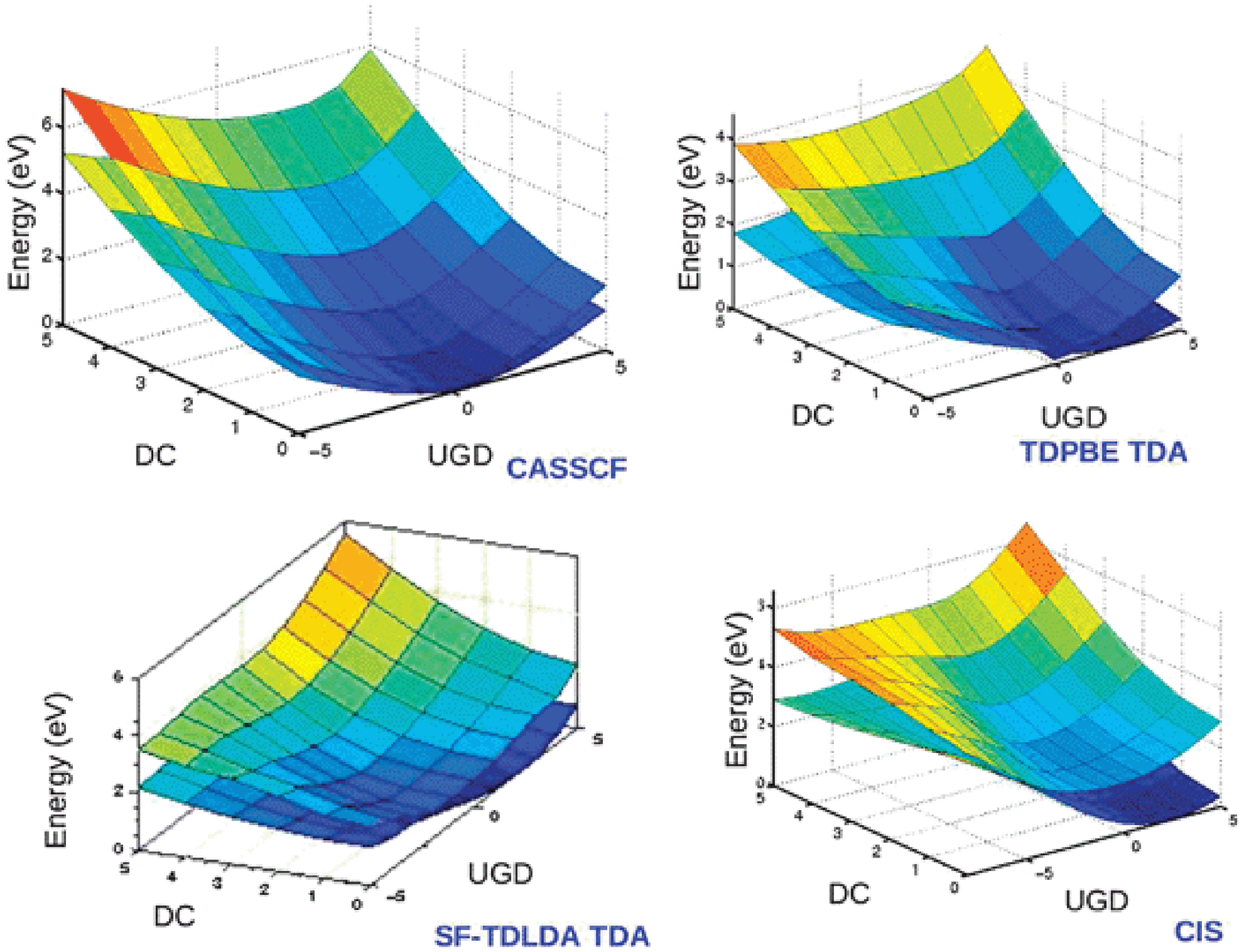}
   \end{center}
   \caption{\label{fig:cx1}
            Comparison of the $S_0$ and $S_1$ potential energy surfaces calculated using different methods for the CASSCF branching
            coordinate space.  
            M.\ Huix-Rotllant, B.\ Natarajan, A.\ Ipatov, C.\ M.\ Wawire, T.\ Deutsch, and M.\ E.\ Casida,
   {\em Phys.\ Chem.\ Chem.\ Phys.} {\bf 12}, 12811 (2010) --- Reproduced by permission of the PCCP
   Owner Societies.
            }
\end{figure}

The oxirane photochemical ring-opening passes through a conical intersection, providing a 
concrete example of a conical intersection to
study with TDDFT.  We now return to the study by Levine, Ko, Quenneville, and Martinez of conical
intersections
using conventional TDDFT \cite{LKQM06} who noted that strict conical intersections are forbidden by the TDDFT adiabatic
approximation for the simple reason that there is no coupling matrix element [Eq.~\ref{eq:pathway_4.1_2}]
to zero out between the ground and excited states.  Figure~\ref{fig:cx1} 
shows a CASSCF conical intersection close to the oxirane photochemical funnel.  Also shown are the 
TDA TDDFT surfaces calculated with the same CASSCF branching coordinates.  Interestingly the CASSCF
and TDDFT conical intersections look remarkably similar.  However closer examination shows that 
the TDDFT ``conical intersection'' is actually
two {\em intersecting} cones rather than a true conical intersection, confirming the observation of Levine {\em et al}.
This was analyzed in detail in Ref.~\cite{TTR+08} where it was concluded that 
the problem is that we are encountering effective noninteracting $v$-representability. 
True noninteracting $v$-representability means that there is no noninteracting system whose ground state
gives the ground state density of the interacting system.  This only means that there is some excited state
of the noninteracting system with integer occupation number which gives the ground state density of the
interacting system.  What we call effective noninteracting $v$-representability is when the LUMO
falls below the HOMO (or, in the language of solid-state physics, there is a ``hole below the Fermi level'').
This is exactly what frequently happens in the funnel region.  


\begin{figure}
  \begin{center}
    \includegraphics[angle=0,width=0.5\textwidth]{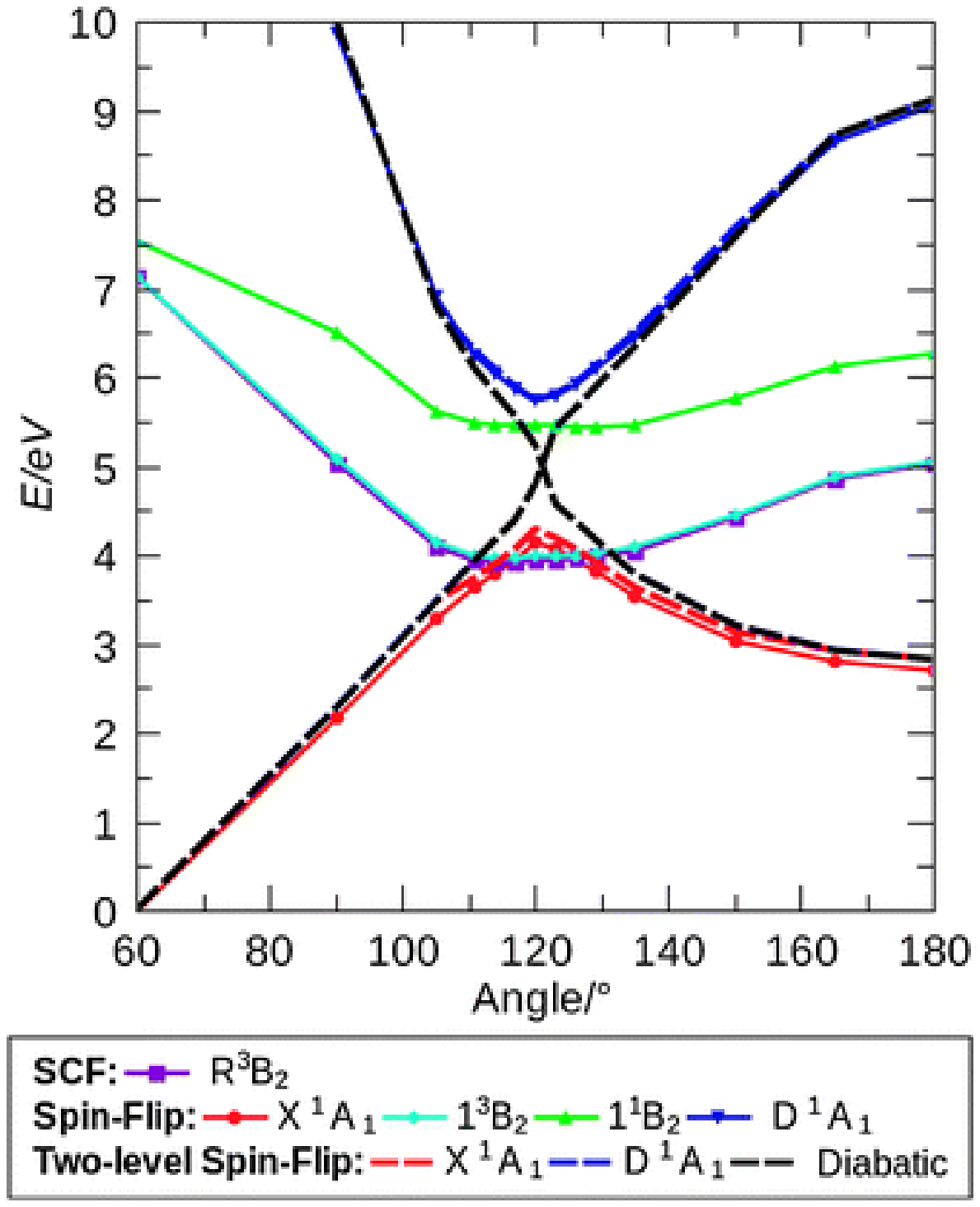}
  \end{center}
  \caption{\label{fig:C2v_2}
  $C_{2v}$ potential energy curves: full calculation (solid lines),
   two-orbital model (dashed lines).
   M.\ Huix-Rotllant, B.\ Natarajan, A.\ Ipatov, C.\ M.\ Wawire, T.\ Deutsch, and M.\ E.\ Casida,
   {\em Phys.\ Chem.\ Chem.\ Phys.} {\bf 12}, 12811 (2010) --- Reproduced by permission of the PCCP
   Owner Societies.
    }
\end{figure}

Spin-flip (SF) TDDFT \cite{SK03,SHK03,WZ04} offers one way to circumvent some of the problems of effective noninteracting $v$-representability in funnel regions.
This is because we can start from the lowest triplet state which has fewer effective noninteracting $v$-representability problems and then 
use SFs to obtain both the ground state and a doubly-excited state.  
Analytic derivatives are now available for some types of SF-TDDFT \cite{SMZ10}.
Figure \ref{fig:C2v_2}
shows that SF-TDDFT works fairly well for treating
the avoided crossing in the $C_{2v}$ ring-opening pathway of oxirane. 
Minezawa and Gordon also used SF-TDDFT to locate a conical intersection in ethylene \cite{MG09}.  However Huix-Rotllant, Natrajan,
Ipatov, Wawire, Deutsch, and Casida found that, although SF-TDDFT does give a true conical intersection in the photochemical
ring opening of oxirane, the funnel is significantly shifted from the position of the CASSCF conical intersection \cite{HNI+10}.
The reason is that the key funnel region involves an active space of over two orbitals which is too large
to be described accurately by SF-TDDFT.

There are other ways to try to build two- and higher-excitation character into a DFT treatment of excited
states.  Let us mention here only multireference configuration interaction (MRCI)/DFT \cite{GW99}, constrained
density functional theory-configuration interaction (CDFT-CI) \cite{WCV07}, and mixed
TDDFT/many-body theory methods based upon the Bethe-Salpeter equation \cite{RSB+09} or the related polarization
propagator approach \cite{C05,HC10} or the simpler dressed TDDFT \cite{MZCB04,CZMB04,GB09,MW09,MMWA10,HIRC11} 
approach. 

All of these may have the potential to improve the DFT-based description of funnel regions in photochemical 
reactions.  Here however we must be aware that we may be in the process of building a theory which is less 
automatic and requires the high amount of user intervention typical of present day CASSCF calculations.  
This is certainly the case with CDFT-CI which has already achieved some success in describing conical 
intersections \cite{KV10}.

\section{Perspectives}
\label{sec:conclude_4.1}

Perhaps the essence of dynamics can be captured in a simple sentence:
``You should from whence you are coming and to where you are going.''
Of course this rather deterministic statement must be interpreted differently in classical
and quantum mechanics.  Here however we would like to think about its meaning in terms of
the development of DFT for applications in photoprocesses.  Theoretical developments
in this area have been remarkable in recent years, opening up the possibility for a more
detailed understanding of femtosecond (and now also attosecond) spectroscopy.
In this chapter we have tried to discuss the past, the present, and a bit of the future.

The past treated here has been the vast area of static investigation and dynamic simulations of photophysical 
and photochemical processes.  We have first described more traditonal wave-function techniques.
We have also mentionned and made appropriate references to important work on early DFT
work involving Ehrenfest TDDFT and restricted open-shell Kohn-Sham  DFT dynamics.
Our emphasis has been on photochemical processes involving several PESs, partly because
of our own personal experiences, but also because photo{\em chemical} processes start out
as photo{\em physical} processes in the Franck-Condon region and then rapidly become more
complicated to handle.

The present-day status of DFT photodynamics is perhaps best represented by the recent availability of 
mixed TDDFT and TDDFTB/classical surface-hopping dynamics codes as well as serious efforts to 
investigate and improve the quality of the TDDFT description of photochemical funnel
regions.  First applications have already shown the utility of this theory and we feel
sure that other applications will follow as programs are made broadly available to
computational scientists.  Finally we have ended the last section with some speculations 
about the future concerning the need for explicit double- and higher-excitations to correctly 
describe funnel regions.

As expected, we could not treat everything of relevance to the chapter title.
Roi Baer's recent work indicating that Berry phase information
is somehow included in the ground-state charge density is most intriguing \cite{B10}.  Also
on-going work on multicomponent DFT capable of treating electrons and nuclei on more or less the
same footing \cite{KG01,KLG08} would seem to open up new possibililties for developing useful
non-Born-Oppenheimer approximations within a DFT framework.
We are sure that still other potentially relevant work has been unfortunately omitted either
because of space limitations or for other reasons.

Do we know where this field is going?  
Certainly non-Born-Oppenheimer photodynamics using some form of DFT is currently a hot and rapidly
evolving area.  Exactly what lies in store may not yet be clear, but what we do know is that
we are going to have fun getting there!

\section*{Acknowledgments}
B.\ N.\ would like to acknowledge a scholarship from the {\em Fondation Nanoscience}.
M.\ E.\ C.\ would like to acknowledge profitable especially discussions with Lorenz Cederbaum, 
Felipe Cordova, Imgard Frank, Todd Martinez, Mike Robb, Enrico Tapavicza, and Ivano Tavernelli.
This work has been carried out in the context of the French Rh\^one-Alpes
{\em R\'eseau th\'ematique de recherche avanc\'ee (RTRA): Nanosciences aux limites de la nano\'electronique}
and the Rh\^one-Alpes Associated Node of the European Theoretical Spectroscopy Facility (ETSF).




\end{document}